# Interrelations among critical current density, irreversibility field and pseudogap in hole doped high-$T_c$ cuprates

S. H. Naqib*, R. S. Islam

Department of Physics, University of Rajshahi, Rajshahi 6205, Bangladesh

*Corresponding author; Email: salehnaqib@yahoo.com

**Abstract**

The effects of hole content (p) and oxygen deficiency (δ) on the zero-field critical current density, $J_{c0}$, were investigated for high-quality c-axis oriented $Y_{1-x}Ca_xBa_2Cu_3O_{7-\delta}$ (x = 0, 0.05, 0.10, and 0.20) thin films. Low temperature critical current density of these films above the optimum doping were found to be high and were primarily determined by the hole concentration, reaching a maximum at p ∼ 0.185 ± 0.005, irrespective of the level of oxygen deficiency. This implies that oxygen disorder plays only a secondary role and the intrinsic $J_{c0}$ is primarily governed by the carrier concentration in the copper oxide planes. Further support in favor of this was found from the analysis of the in-plane resistive transitions of c-axis oriented crystalline thin films of $YBa_2Cu_3O_{7-\delta}$ (YBCO) under magnetic fields (H) applied along the c-direction, over a wide range of doped holes. The characteristic magnetic field ($H_0$), linked to the vortex activation energy and the irreversibility field, exhibits similar p-dependence as shown by $J_{c0}(p)$. We have explained these observations in terms of the doping dependent pseudogap (PG) in the low-energy electronic energy density of states. Both the intrinsic critical current density and the irreversibility field depend directly on the superconducting condensation energy, which in turn is largely controlled by the magnitude of the hole concentration dependent PG in the quasiparticle spectral density.

**Keywords:** Copper oxide superconductors; Critical current density; Irreversibility field; Vortex dynamics; Pseudogap

## 1. Introduction

All large scale applications of high-temperature superconductors depend crucially on two particular parameters; the temperature and field dependent critical current density, $J_c(T,H)$ [1 – 3] and temperature dependent irreversibility magnetic field, $H_{irr}(T)$ [4 – 6]. The superconducting critical current density sets the limit of dissipation-less current density and the irreversibility magnetic field sets the limiting field below which flux lines remain effectively pinned. Both $J_c$ and $H_{irr}$ are influenced by a number of superconducting state and structural parameters [1 – 9]. The factors affecting $J_c$ and $H_{irr}$ can be categorized as intrinsic and extrinsic. The intrinsic factors are the superconducting condensation energy, superfluid density, coherence length, degree of structural and electronic anisotropy etc. On the other hand, extrinsic factors are the grain boundaries, oxygen deficiency, impurity atoms, twins and other tailored defects (e.g., those due to high energy particle irradiation). It is a matter of interest to understand the relative role of extrinsic and intrinsic factors in determining the overall critical current density and irreversibility magnetic field of high-$T_c$ cuprates.

It has been shown in melt-processed YBCO that the angle between the grain boundaries is an important issue. Both $J_c$ and $H_{irr}$ are enhanced in case of low-angle grain boundaries [10, 11]. The conductivity of the grain boundary can also play a significant role [12]. Introduction of impurity based pinning centers of certain type which are not detrimental to superconductivity can also elevate $J_c$ and $H_{irr}$ to a certain extent [13]. On the other hand, a number of prior studies have shown clearly that intrinsic superconducting state parameters related to the depairing critical current density play the major role in maximizing both $J_c$ and $H_{irr}$ [7 – 9, 14].

The fundamental physics of high-$T_c$ cuprates are controlled by the number of doped holes in the $CuO_2$ plane [14 – 17]. Superconductivity emerges from an antiferromagnetically ordered insulating phase upon hole doping beyond a certain level [14 – 17]. Addition of doped holes (p) in the $CuO_2$ plane destroys magnetic order and transforms the compound into a *bad metal* and superconductivity is found with varying critical temperatures in the doping range $0.05 < p < 0.28$. The transition temperature is a maximum for $p \sim 0.16$, known as the optimum hole content separating the underdoped ($p < 0.16$) and overdoped ($p > 0.16$) regions in the $T_c(p)$ diagram [18, 19]. Beside superconductivity, almost all the normal state properties of cuprates are strong functions of p. The normal state properties in the underdoped region are anomalous in the sense that ordinary Fermi-liquid theory seems inadequate to describe the experimental observations [20 – 22]. It is believed that the root cause of these anomalous normal state behaviors is the pseudogap (PG) in the low-energy quasiparticle (QP) energy spectrum [20 – 23]. The PG also affects all the superconducting state parameters, including those related to the depairing critical current density [7 – 9, 14]. The electronic phase diagram of hole doped cuprates are quite generic as far as the hole content dependent superconducting transition temperature and the PG energy/temperature scale, $T^*$, are concerned [20 – 23]. The almost parabolic $T_c(p)$ behavior is found in all hole doped cuprate families irrespective of the maximum superconducting $T_c$ which shows a wide variation in the range from 35 K to 164 K [14, 18, 23, 24]. $T^*(p)$ on the other hand falls almost linearly with increasing hole content in the underdoped region and appears to vanish at p ~ 0.19 under the superconducting dome [23, 25]. The evolution of $T^*(p)$ is not completely resolved though [20 – 22]. Unlike the wide variation of the maximum $T_c$ for different families of cuprates, the T*(p) values are almost identical for all the hole doped cuprate superconductors.

In this communications, we wish to summarize results obtained from our previous investigations to present a unified picture regarding the role of the PG on superconducting critical current density and the irreversibility magnetic field [26 – 28]. We will focus on high-quality thin films of pure and Ca substituted YBCO compounds. Results of critical current measurements from the magnetization and resistivity measurements under magnetic fields will be considered at different levels of doped holes in the Ca-YBCO superconductors.

## 2. Experimental samples, measurements, and discussions

Very high-quality c-axis oriented thin films of $Y_{1-x}Ca_xBa_2Cu_3O_{7-\delta}$ (Ca-YBCO) were grown from high-density single-phase sintered targets using the method of pulsed laser deposition (PLD). The films were deposited on (001) $SrTiO_3$ substrates. All the films were characterized using the X-ray diffraction (XRD), atomic force microscopy (AFM), ab-plane room-temperature thermopower, $S_{ab}[290K]$, and in-plane resistivity $\rho_{ab}(T)$, measurements. The films were phase-pure and showed high-degree of c-axis orientation. Thicknesses of the films were in the range (2700 ± 300) Å. Room-temperature thermopower was used to calculate the planar hole content following the work by Tallon et al. [29]. The level of oxygen deficiency was estimated from an earlier study where the relation between δ and p were established [30]. The ab-plane resistivity gave information about the impurity content and the quality of the grain-boundaries of the films. All our samples had low values of $\rho_{ab}(300\ K)$ and the extrapolated zero temperature resistivity; the residual resistivity $\rho_{ab}(0\ K)$ [31]. Further details of the thin film fabrication and their characterizations can be found in Refs. [30, 31]. The hole content of the Ca-YBCO thin films were controlled in two different ways - by controlling the level of oxygen deficiency (δ) in the $CuO_{1-\delta}$ chains via annealing at different temperatures under flowing oxygen [30, 31] and by substituting divalent Ca in place of trivalent Y. Pure YBCO with full oxygenation ($\delta = 0$) is slightly overdoped (p ~ 0.18). Ca substitution gives access to the deeply overdoped region.

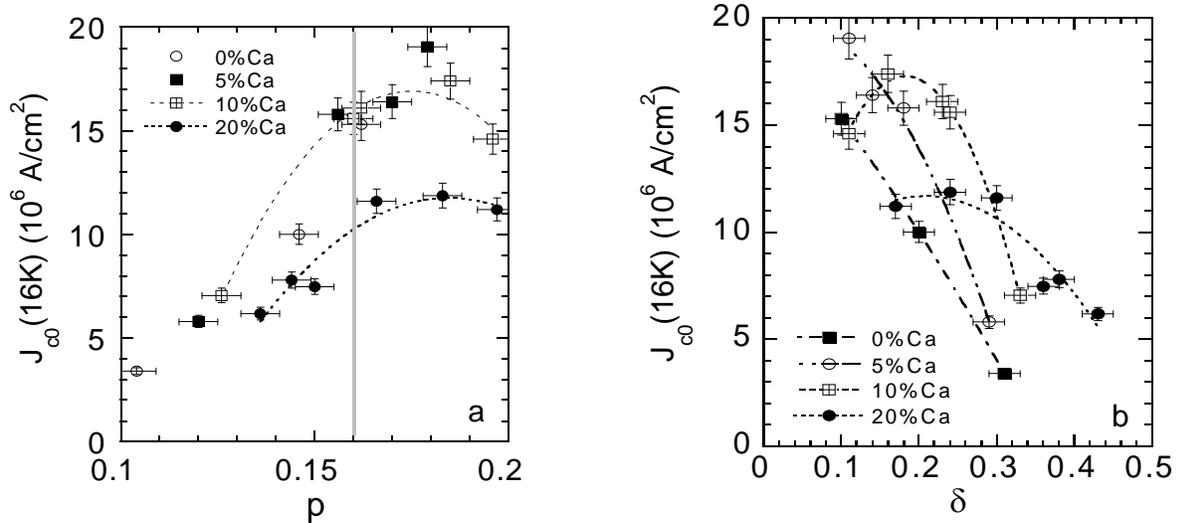

Figure 1. (a) $J_{c0}$(16K) versus p results for $Y_{1-x}Ca_xBa_2Cu_3O_{7-\delta}$ thin films. The dashed curves show the trends of $J_{c0}$(16K) as a function of p for the 10% and 20% Ca-YBCO. The vertical line shows the position of the optimum doing level where the $T_c$ is maximum and (b) $J_{c0}$(16K) versus $\delta$ for the Ca-YBCO thin films. The dashed curves show the trends of $J_{c0}$(16K) as a function of $\delta$ (taken from Ref. 8).

We show the low-temperature zero-field critical current density of Ca-YBCO thin films in Figs. 1. The zero-field $J_c$ was obtained from the magnetization loops (M-H curves) measured at different fixed temperatures using modified critical state formalism [32]. The magnetic field was applied along the c-direction of the Ca-YBCO thin films. Under this field configuration, the supercurrent flowed exclusively in the $CuO_2$ planes. Details regarding the magnetization measurements and calculation of the critical current density can be found elsewhere [8]. It is interesting to note that, irrespective of Ca content (x), low-temperature $J_c$ is maximized for p ~ 0.185 in the overdoped side. This particular hole content is attained with different levels of oxygen deficiency in the Cu-O chains in the Ca-YBCO thin films having different values of x (Figs. 1a and 1b). This is a clear demonstration that maximum $J_c$ is determined by the hole content in the $CuO_2$ planes and not by the level of oxygen vacancy in the Cu-O chains. This also implies that the oxygen vacancy/disorder does not act as strong pinning centers for trapped magnetic flux lines. An important feature of Fig. 1a is the observation that the hole content at which $J_c$ is maximized is not equal to the optimum doping where $T_c$ is at its largest. The decrease of $J_c$ in the overdoped side for $p > 0.19$ is due to the growth of quantum disordered metallic state that weakens superconducting correlations.

To explore the vortex dynamics, the nature of resistive transitions under magnetic field was investigated for YBCO thin films. The broadening of superconducting (resistive) transition under magnetic fields is directly related to the motion of the magnetic flux lines [33-35]. Representative plots showing the $\rho_{ab}(T, H)$ data of pure YBCO with different hole concentrations is shown in Figs. 2. Details regarding the resistivity measurements with magnetic fields along the c-direction can be found in Ref. [26].

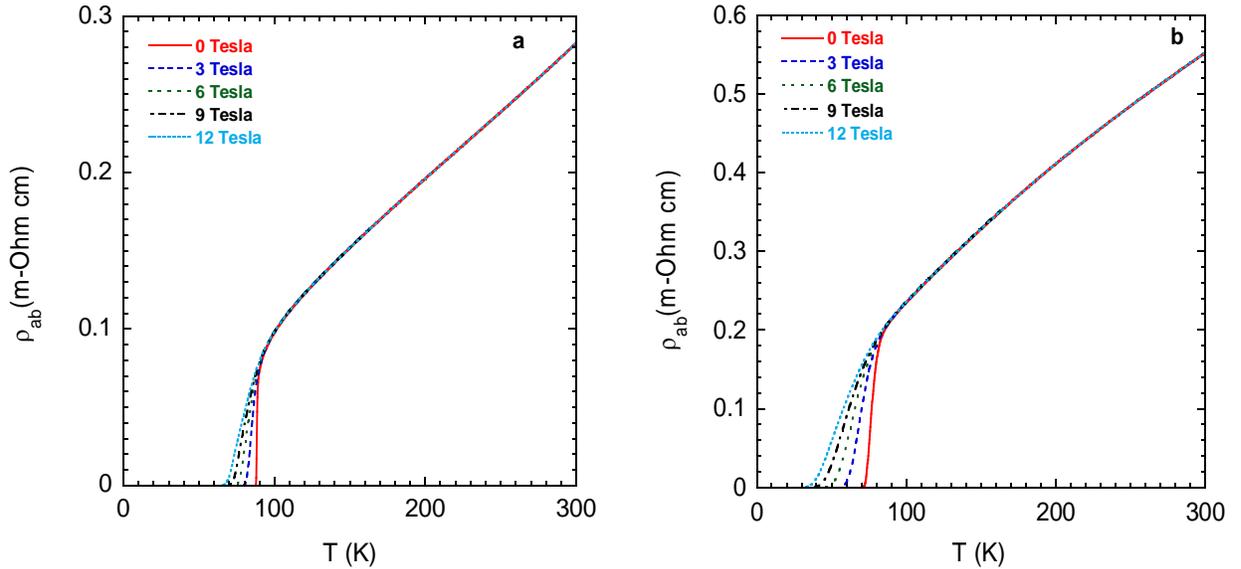

Figure 2: Representative $\rho_{ab}(T, H)$ of (a) YBCO with $p = 0.15$ and (b) $p = 0.12$ thin films under different magnetic fields applied along the crystallographic *c*-direction (taken from Ref. 26).

From the scaling analysis of the $\rho_{ab}(T, H)$ data, useful information regarding the temperature and magnetic field dependent flux pinning energy can be extracted [26]. The formalism is described in detail elsewhere [26, 33-35]. The scaling analysis gives a value of the characteristic magnetic field $H_0$ which is closely related to the activation energy of the magnetic vortices and irreversibility magnetic field. We show the hole content dependent $H_0$ for YBCO thin films in Fig. 3.

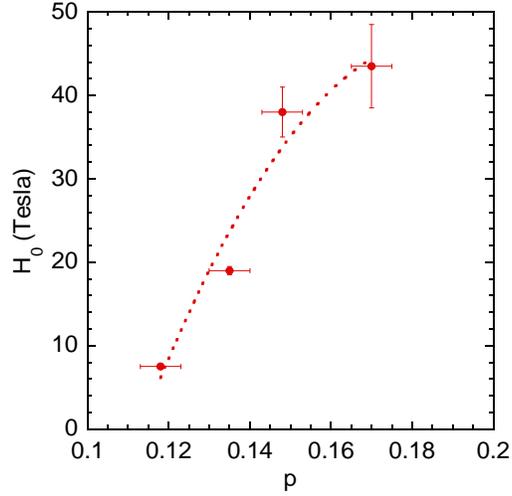

Figure 3: Variation of the characteristic magnetic field $H_0$ with hole concentration p in the $CuO_2$ planes of YBCO. The dashed line shows the overall trend.

Since both critical current density and $H_0$ depend directly on the vortex pinning/activation energy, the similarity in the trends of $J_{c0}(p)$ and $H_0(p)$ is not surprising.

The characteristic PG energy scale, $T^*$, can be estimated from the $\rho_{ab}(T)$ data. $T^*$ is located at the onset of downward deviation of $\rho_{ab}(T)$ from its high-temperature linear behavior in the normal state at temperatures much above the $T_c$. An example of this procedure is given in Fig. 4 below [30].

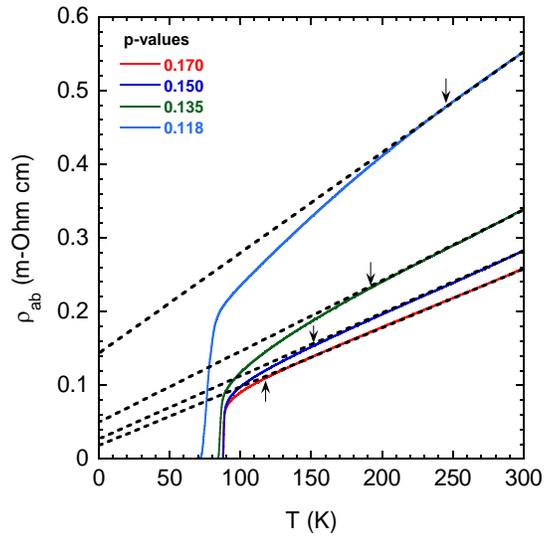

Figure 4: $\rho_{ab}(T)$ plots for YBCO thin films with different p. The arrows mark the PG temperature, $T^*$.

The PG induced downturn in the resistivity can be understood qualitatively in the following terms. In the presence of pseudogap the available low-energy quasiparticle states near the chemical potential are depleted. The degree of depletion is higher for cuprates having higher $T^*$. This depletion in the electronic energy density of states reduces the QP scattering rate, thereby reducing the resistivity at high-temperatures much above the superconducting transition. Fig. 5 shows a plot of $T^*(p)$ against $H_0(p)$. A clear anti-correlation is observed; $H_0(p)$ falls linearly as $T^*(p)$ increases due to progressive underdoping.

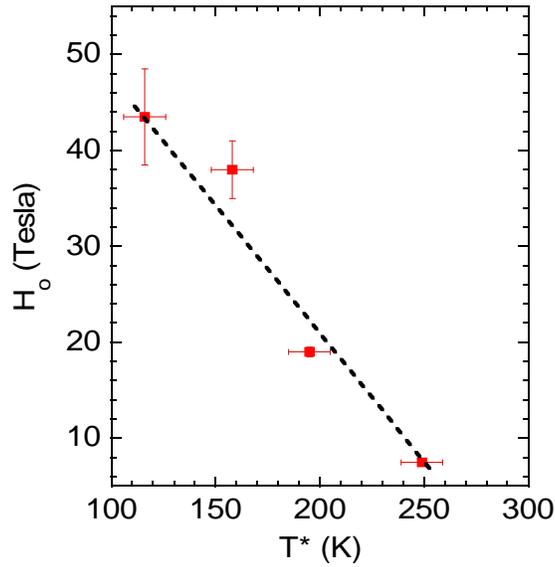

Figure 5: $H_0(p)$ versus $T^*(p)$. An anti-correlation is observed. The thick dashed straight line is drawn only as a guide for the eye.

At this point the question that arises is how the doping dependent magnitude of PG affects the doping dependent superconducting critical current density and the irreversibility magnetic field. We believe that the answer lies on the effect of the pseudogap on the superfluid density. This particular parameter controls the depairing critical current density to a large extent. It is understood that the supercurrent flows in a superconductor due to the gradient of the *phase angle* (measured through $\nabla\theta$) of the order parameter. Value of the gradient depends on the *phase stiffness* of the order parameter. The phase stiffness, on the other hand, varies linearly with the superfluid density [36]. Therefore, from the intrinsic considerations, depairing critical current density is expected to vary systematically with the superfluid density or the superpair density [37].

The role of the pseudogap can be envisaged from the following argument. The magnetic flux lines are pinned at locations (pinning sites) where the superconducting order parameter is partially or almost completely suppressed. As a consequence the pinning energy of the vortex core reveals itself as the energy barrier to the dissipative movement of the flux line and gives us a clear measure of the activation energy $U_a$ for the pinned flux line [38]. It is this activation energy that sets the values of $J_c$ and the irreversibility magnetic field [26, 38] of a superconductor. Using a very simple heuristic scaling, Yeshurun and Malozemoff [39] and Tinkham [40] demonstrated that $U_a \sim H_c^2$, where $H_c$ is the thermodynamic critical magnetic field. On the other hand, the superconducting condensation energy can also be expressed as $U_0 \sim H_c^2$. Therefore, one gets $U_a \sim U_0 \sim H_c^2$ [26, 38]. Furthermore, the superconducting condensation energy can also be found from: $U_0 = <N(E_F)>\Delta_{sc}^2$, where $<N(E_F)>$ is the thermal average of the normal state electronic energy density of states over an energy width $\sim\pm\ 3k_BT_c$ centered at the Fermi level and $\Delta_{sc}$ is the amplitude of the superconducting energy gap. This relation for $U_0$ establishes a direct link among vortex dynamics with an activation energy scale $U_a$, and the superpair density $\rho_s(p)$, given by $\rho_s(p) = <N(E_F)>\Delta_{sc}$. The pseudogap depletes the $N(E_F)$ and reduces the superpair density in the superconducting state. As a result the superconducting condensation energy decreases. Considering a hole content dependent PG at the Fermi level, we have checked the assumption $U_0 = <N(E_F)>\Delta_{sc}^2 \sim U_a$ and its relevance to the critical current density in details in a recent work [37] and consistent results have been found. To explore the correspondence between $J_c$ and $U_0$, we show the hole content dependent variations of both these parameters in Fig. 6. In this figure superconducting condensation energy was taken from the analysis of specific heat results of $Y_{0.80}Ca_{0.20}Ba_2Cu_3O_{7-\delta}$ [41] sintered compounds. The condensation energy does not depend on the crystalline state of superconductors. Fig. 6 shows a strong correspondence between $J_c(p)$ and $U_0(p)$.

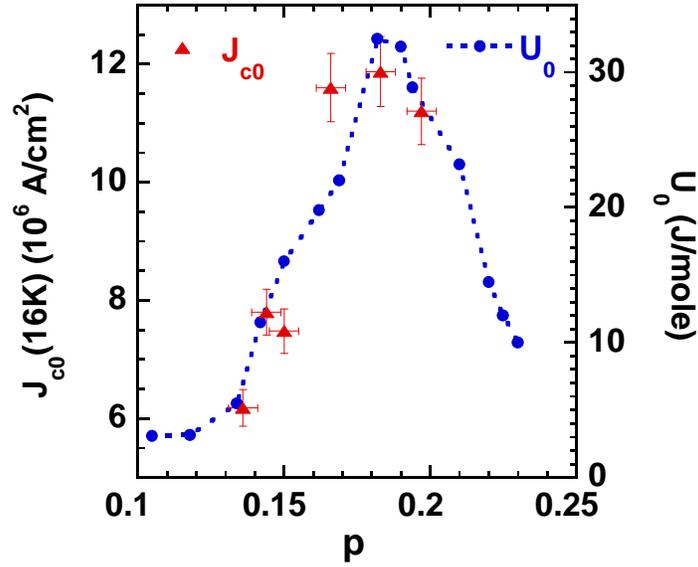

Figure 6: Variation of low-T zero-field critical current density of $Y_{0.80}Ca_{0.20}Ba_2Cu_3O_{7-\delta}$ thin films and superconducting condensation energy of $Y_{0.80}Ca_{0.20}Ba_2Cu_3O_{7-\delta}$ sintered samples with doped hole concentration in the $CuO_2$ planes.

## 3. Conclusions

From the analysis of charge transport and magnetization results we found that $J_c$ peaks at $p \sim 0.185$, irrespective of the values of $\delta$ and $x$ in $Y_{1-x}Ca_xBa_2Cu_3O_{7-\delta}$ thin films. The role of oxygen disorder is secondary, $J_c$ is mainly determined by the hole content in the $CuO_2$ planes. The hole content at which $J_c$ is maximized is very close to the one where the pseudogap energy/temperature scale goes to zero. Like superconducting critical current density, the characteristic magnetic field $H_0$, that primarily determines the pinning strength, decreases sharply with decreasing p in the underdoped side. It appears that both $J_c$ and $H_0$ are controlled by the superconducting condensation energy, and therefore, is linked directly to the pseudogap in the quasiparticle energy spectrum.

## Dedication

The corresponding author dedicates this work to the loving memory of his father, Professor A. K. M. Mohiuddin, who passed away early this year.

## Declaration of interest

The authors declare that they have no known competing financial interests or personal relationships that could have appeared to influence the work reported in this paper.

## Data availability

The data sets generated and/or analyzed in this study are available from the corresponding author on reasonable request.

## CRediT author statement

**S. H. Naqib:** Conceptualization, Measurements, Supervision, Analysis, Writing- Reviewing and Editing. **R. S. Islam:** Measurements, Writing- Reviewing and Editing.